\documentclass[12pt]{iopart}

\usepackage{iopams}
\usepackage{amstext}
\usepackage{graphicx}

\begin{document}

\title[]{Hadronization via Recombination}

\author{Kang Seog Lee$^a$\footnote{kslee@chonnam.ac.kr},  S. Bass$^b$, B. M\"uller$^b$ ,and C. Nonaka$^c$ }
\address{$^a$Department of Physics, Chonnam National University, Korea}
\address{$^b$Department of Physics, Duke University, USA}
\address{$^c$Department of Physics, Nagoya University, Japan}

\begin{abstract}

The recombination model as a model for hadronization from a quark-gluon plasma has been recently revived since it has advantages in explaining several important features of the final state produced in heavy-ion collisions at RHIC, such as the constituent quark number scaling of the elliptic coefficient versus the transverse energy of identified hadrons, the bending shape of the $p_T$ spectrum of hadrons near 5 GeV/c, and the measured large value of baryon to meson ratio(of the order of unity) in the same $p_T$ range. We have developed a dynamic simulation model of heavy-ion collisions in which a quark-gluon plasma, starting from a certain initial condition, evolves hydrodynamically until it reaches the phase boundary, and then hadronizes by valence quark recombination. Rescattering after hadronization is described by UrQMD. We discuss some details of the model and report first, preliminary results.
\end{abstract}



\section{Introduction }

The quark recombination model of hadron formation, which was originally proposed in 1970s\cite{das} and studied actively in 1980s\cite{biro}, has been revived\cite{hwa,fries,greco,lin,voloshin} as it naturally explains some of the features of the produced hadrons in relativistic heavy-ion collisions. The hydrodynamic analysis of the measured elliptic flow coefficient of hadrons, which signals the collective behavior of the produced system, could explain the magnitude and the ordering of the magnitude of the coefficient $v_2$ of various hadrons and thus showed that the system produced during the heavy-ion collisions behaves as a strongly interacting, nearly perfect fluid.  If one plot the elliptic coefficient per constituent quarks, $v_2/n$ versus the transverse kinetic energy per constituent quarks, $E_T /n$, then the data for various hadrons converge into a single line, suggesting that the collectivity is the property not of the hadrons but also of the constituent (valence) quarks. A natural explanation is that the hadrons are produced by recombining the quarks which are collectively moving in the quark-gluon plasma.

A second set of data supporting the recombination of quarks to form hadrons are the transverse momentum spectra of hadrons, which shows bending behavior near $p_T \sim 5$ GeV. Hadrons with small momentum ($p_T \sim 2$ GeV) are thermal ones and large momentum hadrons are those from jets. Fragmentation of jets may produce hadrons with $p_T \gtrsim 5$ GeV while the recombination of quarks with thermal distribution in the quark-gluon plasma can produce the hadrons with $p_T \lesssim 5$. The two mechanisms combined have shown to fit the hadron spectra in the mid-$p_T$ range well. Another experimental observation supporting the recombination model is the large baryon-to-meson ratio, {\it e.g.} $R(p/\pi) \sim 1$, near $p_T \sim 2$ GeV. If the hadrons are formed by recombining quarks with a thermal distribution, the probabilities of forming mesons or baryons are determined only by the probabilities to find quarks, and the ratio $R(p/\pi) \sim 1$ can be naturally explained.

The recombination calculations reported so far were mostly done in the framework of schematic evolution models. The only dynamical calculation is that by Ko {\it et. al.}: The initial partonic matter evolves through the parton cascade and at the phase boundary hadrons are formed by recombination. Thus fixing the initial state just after the the nuclear impact, they can simulate the main experimental features described above. Here we report a dynamic simulation method describing the heavy-ion collisions where the evolution of the quark-gluon plasma is described by ideal hydrodynamics up to the phase boundary. At the phase boundary, hadronization is simulated by the recombination mechanism and the produced hadrons may or may not further rescatter until thermal freeze-out. In our implementation of this model, quark-gluon evolution is described by the hydrodynamic equations. For the hydrodynamic cells at the phase boundary, hadronization occurs via recombination model and thus the Monte Carlo program creates hadrons with their definite position and momentum at each cell whose expansion velocity is explicitly known. The hadrons may further undergo hadronic rescattering and this process is described by UrQMD.

\section{Dynamic Recombination Model}


We use the three-dimensional ideal relativistic hydrodynamic program by C.~Nonaka \cite{nonaka} which uses comoving (Lagrangian) grid points rather than the commonly used fixed(in some coordinate system) Eulerian grid. The advantage of the Lagrangian grid for the calculation the hypersurface integral of the type appearing in the Cooper-Frye formula will be explained below.

Each quark-gluon plasma fluid cell at $T_c$ is specified by the temperature, $T_c$, the baryon chemical potential, $\mu_B$, position,  $(\tau, x, y, \eta)$, and the spatial length, $(d\tau, dx,dy, d\eta)$,  where $\tau=\sqrt{t^2-z^2}$ is the proper time and $\eta={\rm artanh}(z/t)$ is the space-time rapidity. The cell is moving with its 4-velocity $u^\mu = \gamma (1,\vec{v})$. The quark distribution in the cell at hadronization is given as
\begin{equation}
 w_a (p^\mu )= \gamma_a \exp({-p^\mu }u_\mu /T ).
\end{equation}

The invariant cross sections for recombined hadrons are given as \cite{fries2}:
\begin{equation}
E\frac{dN_{M_i} }{dP^3} = \frac{C_i}{(2\pi)^3} g_M (P)
\end{equation}
\begin{equation}
g_M (P) = \int_\Sigma P^\mu d\sigma_\mu \int d^3k\, |\phi_M (x)|^2 w_a(k) w_b(P-k)
\end{equation}
for mesons, and
\begin{equation}
E\frac{dN_{B_i} }{dP^3} = \frac{C_i}{(2\pi)^3} g_B (P)
\end{equation}
\begin{equation}
g_B (P) = \int_\Sigma P^\mu d\sigma_\mu \int d^3k_1  d^3k_2\,
|\phi_B (x_1 ,x_2)|^2 w_a (k_1) w_b (k_2) w_c (P-k_1 -k_2 )
\end{equation}
for baryons, where $\Sigma$ denotes the integration over the hadronization hypersurface.
Inserting the quark distribution into the above equations, one gets
\begin{equation}
g_M (P) =
 \int p^\mu d\sigma_\mu  \exp(\vec{P}\cdot\vec{u}/T)\int
d^3 k
\exp(-\gamma (E_a +E_b)/T) |\phi_M (k)|^2
\end{equation}
for mesons, where  $E_a$ and $E_b$ are the energies of the quark and antiquark, respectively. Similarly for baryons:
\begin{eqnarray}
g_B (P) &=& \int p^\mu d\sigma_\mu \exp(\vec{P}\cdot\vec{u}/T) \int d^3 k_1 d^3 k_2 |\phi_B (k_1 ,k_2)|^2
\nonumber \\
& & \times\exp(-\gamma (E_a+E_b+E_c)/T) .
\end{eqnarray}

Since the wave functions for hadrons are best known in the light-front frame, it is convenient to use the light-cone variable, $x$, which denotes the longitudinal momentum fraction carried by the quark in the infinite momentum frame. If one neglects the transverse momentum of quarks relative to the hadron momentum $P$, the equations become simplified. When the wave functions for hadrons are known, one can explicitly decompose the quark momentum into the longitudinal transverse components, $\vec{k_i} = x \vec{P} + \vec{k_{i,\perp}}$, and evaluate the integrals exactly. In the above equations, integration over the hadronization hypersurface $\sigma_\mu$ is done numerically. The hypersurface areas are given by the cell surfaces between adjacent cells, one of which has a temperature $T\geq T_c$ and the other has $T<T_c$. In the case of a first-order phase transition with phase coexistence, also pairs of cells with a different fraction of quark-gluon plasma contribute. A single grid cell may have space-like as well as time-like surfaces, contributing to the hadronization hypersurface. It is known that the hydrodynamic programs using an Eulerian grid sometimes give negative values of the expression $P^\mu d\sigma_\mu$. However, in the hydrodynamic program which uses the Lagrangian coordinate such as developed by C. Nonaka and is used in this calculation, $P^\mu d\sigma_\mu$ is positive definite.

During the hadronization energy and momentum should be conserved, equivalent to energy conservation in the rest frame of a cell. We enforce energy conservation by continuing the creating process of hadrons, using the Monte Carlo method, until all the energy of the cell is exhausted for hadronization. However, since our grid is so fine that the total energy of each individual fluid cell is smaller than the energy required to create a hadron, we add the energies of all fluid cells within a small $\eta$ interval and randomly create hadrons from this ensemble of cells. The Monte Carlo implementation of this scheme works as follows. Consider a fluid cell at $T_c$ in the mixed phase whose degree of hadronization is characterized by the QGP volume fraction $\alpha = V_Q /(V_Q +V_H )$. If $E_i$ is the total energy in the fluid cell $i$ and $\delta\alpha_\tau$ is the negative of the change of $\alpha$ during the evolution in one time step, then $E_i \,\delta\alpha_\tau$ is the energy used for the hadronization during the time interval. Summing the energy for all the cells in a given $\eta$ interval, one finds the total energy to hadronize among the fluid cells in $(\eta,\eta+d\eta)$. Hadronization continues until all the energy has been used to create hadrons.

In the sudden recombination model the probability to create a hadron with certain momentum $P$ depends only on the number and flavor of its constituent quarks but is independent of the hadron mass. In order to implement these constraints, we dived the mesons into four groups with different flavor contents:
\begin{equation}
M1=(\textrm{q} \bar{\textrm{q}}), M2=(\textrm{q} \bar{\textrm{s}}), M3=(\textrm{s} \bar{\textrm{q}}), M4=(\textrm{s} \bar{\textrm{s}}) .
\end{equation}
Similarly, there are a total of eight groups for baryons, four for baryons and four for antibaryons:
\begin{eqnarray}
B1=(\textrm{qqq}) , B2=(\textrm{qqs}), B3=(\textrm{qss}), B4=(\textrm{sss}),
\nonumber \\
AB1=(\bar{\textrm{q}}\bar{\textrm{q}}\bar{\textrm{q}}), AB2=(\bar{\textrm{q}}\bar{\textrm{q}}\bar{\textrm{s}}),
AB3=(\bar{\textrm{q}}\bar{\textrm{s}}\bar{\textrm{s}}), AB4=(\bar{\textrm{s}}\bar{\textrm{s}}\bar{\textrm{s}}) .
\end{eqnarray}
To obtain the probability to hadronize into a hadron with certain flavor content and momentum, one randomly choses a cell among those in the same $\eta$ interval, the flavor group, the momentum and a random number in the interval $[0,1]$. When the random number exceeds the probability to create the hadron given by the recombination model equations, a hadron with certain momentum is created in that fluid cell. This procedure is repeated until all the energy available is exhausted. Finally, in order to determine the hadron species inside a given group (e.g.\ $p$ versus $\Delta^+$, or some other resonance included in the UrQMD model), we assume a relative Boltzmann weight, $d_i e^{-m_i /T}$, where $d_i$ is the degeneracy and $m_i$ the hadron mass.

As a result of hadronization, we create hadrons moving with a thermal velocity spectrum of a fluid cell superimposed on the collective flow velocity of the cell given by the hydrodynamics. Rescattering among those hadrons is then simulated with the help of UrQMD. By running UrQMD code, we may check whether the chemical freeze-out coincides with the time of hadronization, and we can track the cooling of the system until the final thermal freeze-out.

\section{Results and Summary}

\begin{figure}
  \begin{center}
     \begin{tabular}{cc}
       \resizebox{70mm}{!}{\includegraphics[angle=-90]{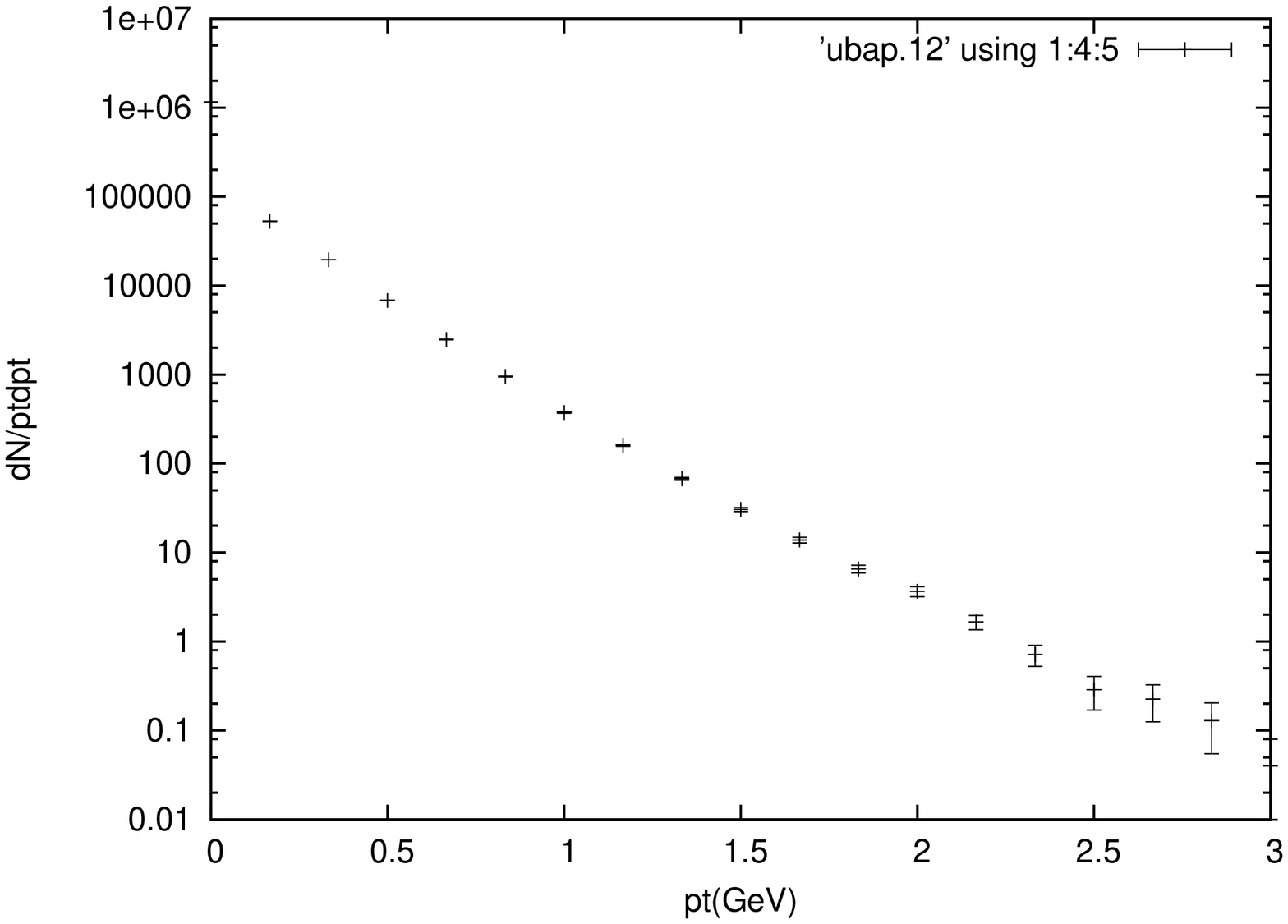}}&
       \resizebox{70mm}{!}{\includegraphics[angle=-90]{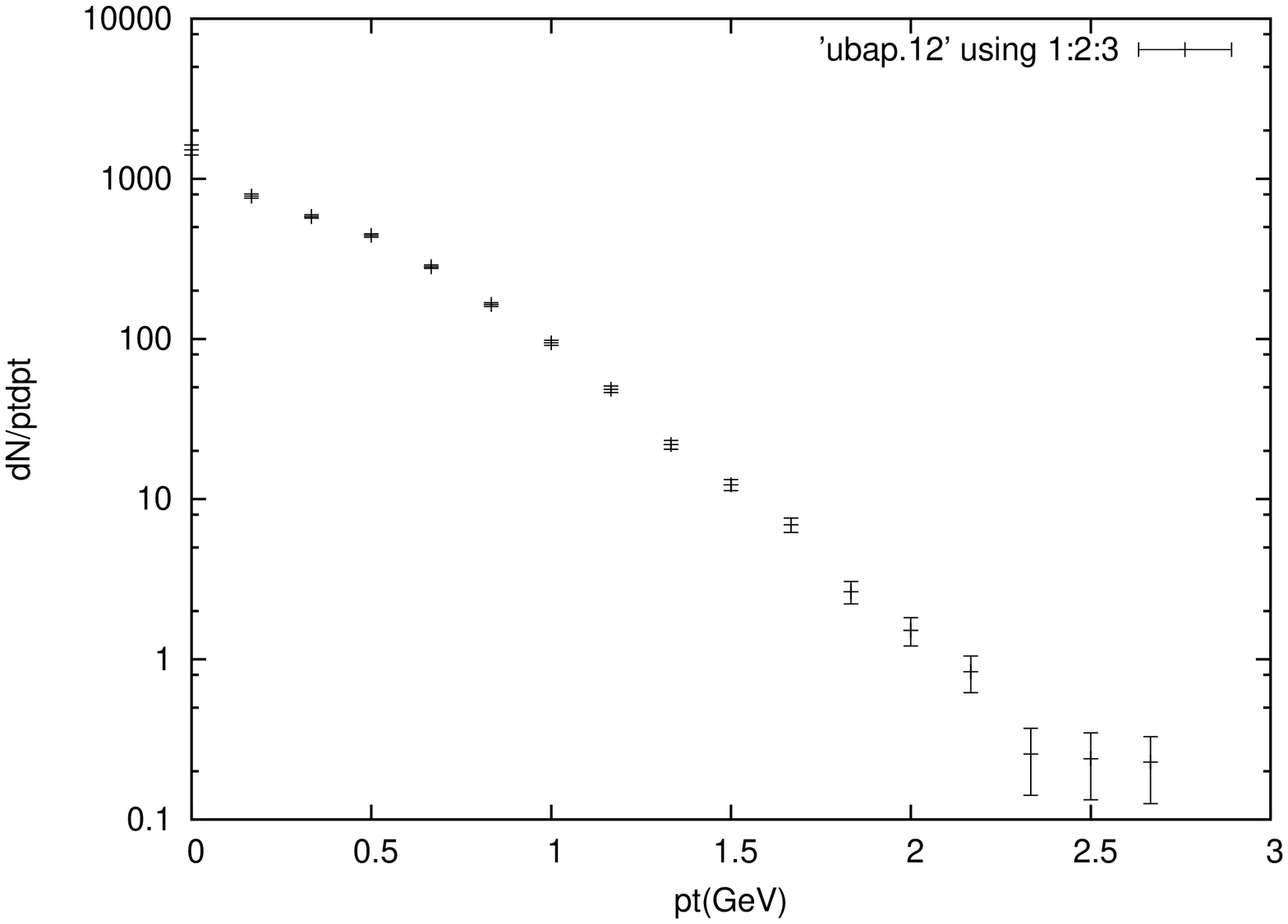}} \\
     \end{tabular}
  \end{center}
   \caption{Transverse momentum spectra of $\pi^-$ (left) and protons (right) hadronized from a QGP with the initial state in Ref.[9], which was used to simulate the central Au+Au collisions at 200 GeV$\cdot$A at RHIC, and experienced the UrQMD afterburner.}
  \label{fig:pt}
\end{figure}

\begin{figure}
  \begin{center}
     \begin{tabular}{cc}
       \resizebox{80mm}{!}{\includegraphics[angle=-90]{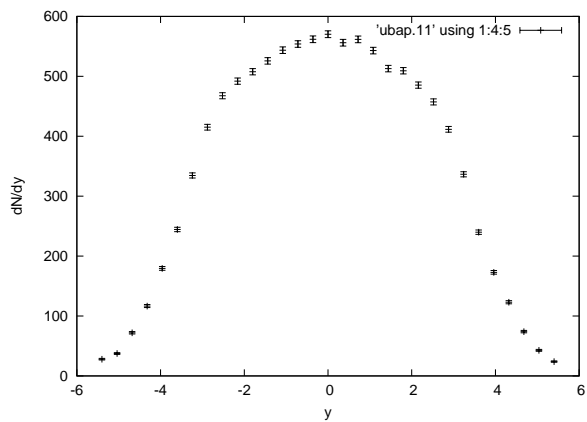}}& 
           \end{tabular}
  \end{center}
   \caption{ $\pi^-$ rapidity spectrum obtained as in Fig. 1}
    \label{fig:dndy}
\end{figure}

In Fig.~\ref{fig:pt} the transverse momentum spectra,  $dN/p_t dp_t$ vs. $p_t$ of pions and protons are shown. Those particles are hadronized from a QGP with an initial state same as in Ref. [9] for the U+U collisions at 200 GeV/A, and experienced the final state interaction by running UrQMD. The results are obtained from 20 events and in order to make the systematic error smaller one has to repeat the simulation for many events.
In Fig.~\ref{fig:dndy} the rapidity spectra of pions are shown.

As emphasized before, hadrons are generated from a QGP according to the sudden recombination model, and rescattering among hadrons is simulated by UrQMD. The creation of hadrons with their position and momentum enables us to``measure'' any physical observables of interest. The only inputs are those of the initial state of the hydrodynamic evolution and the hadronization temperature $T_c$ (or the equation of state). The comparison of the result with the statistical model will be interesting. In the future we hope to calculate of the elliptic flow coefficients of various hadrons in order to study how the flow anisotropy is related to the initial momentum anisotropy of the QGP.

{\em Acknowledgements:} 
This work was supported in parts by Chonnam National University, 2007, the Korea Science and Engineering Foundation(KOSEF) grant funded by the Korea government(MOST) (No. R0120070002107402008), and DOE grant DE-FG02-05ER41367. K. S. Lee wish to thank S. Bass and B. M\"uller for their hospitality during his stay at Duke University.
\bigskip


\end{document}